\documentstyle[11pt,newpasp,twoside,epsf]{article}
\def\edcomment#1{\iffalse\marginpar{\raggedright\sl#1\/}\else\relax\fi} 
\reversemarginpar 
 
\begin{document} 
\title{Intracluster Stellar Population} 
 \author{Magda Arnaboldi} 
\affil{INAF, Astronomical Observatory of Turin, Strada Osservatorio
 20, 10025 Pino Torinese} 
 
\begin{abstract} 
I shall review the latest results for the presence of diffuse light in
nearby clusters, and the evidence of ongoing star formation in an
intracluster Virgo field.  I shall discuss how intracluster planetary
nebulae can be used as excellent tracers of the diffuse stellar
population in nearby clusters. Their number density distribution,
density profile and radial velocity distribution provide
observational constraints to models for cluster formation and
evolution.  The preliminary comparison of the avaliable ICPN samples
with high resolution N-body models of a Virgo-like cluster in a Lambda
CDM cosmology supports ``harassment'' as the most likely mechanism for
the origin of diffuse stellar light in clusters.
\end{abstract}

\section{Introduction: Discovery of diffuse light in clusters} 
Stars are usually observed to form in galaxies (disks, dwarfs and
starbursts). In nearby galaxy clusters, however, a diffuse intracluster
stellar component has been detected from deep imaging and observations
of individual intracluster stars.

Intracluster light (ICL) is potentially of great interest for studies
of galaxy and galaxy cluster evolution. The dynamical evolution of
cluster galaxies involves complex and imperfectly understood processes
such as galactic encounters, cluster accretion, and tidal
stripping. Various studies have suggested that between 10\%
and 50\% of a cluster's total luminosity may be contained in the ICL,
with a strong dependence on the dynamical state of the cluster. The
properties of the ICL may also be sensitive to the distribution of
dark matter in cluster galaxies, as simulations have shown that the
structure of DM halos in galaxies plays a central role in the
formation and evolution of tidal debris (Dubinski et al. 1999).

Recently some progress has been made in the study of intracluster star
light on several fronts. Individual intracluster stars, including planetary
nebulae detected from the ground and red giants detected using HST,
have been discovered in the Virgo cluster. These intracluster (IC)
stars give the promise of studying in detail the kinematics,
metallicity and age of the intracluster stellar population in nearby
galaxy clusters and thereby learning about the origin of this diffuse
stellar component, and the details of the cluster origin.

\section{Intracluster light: an historical perspective} 
The first studies on this subject date back to the work of Zwicky
(1951) on the luminosity function of galaxies in the Coma cluster. In
his work, Zwicky pointed out that vast and irregular {\it swarms
  of stars} existed in the spaces between the standard galaxies in
Coma, and commented whether they could be incorporated into the
distribution function of known galaxy types. Zwicky's efforts were
then followed by photographic surveys for diffuse light in Coma and
other rich clusters in the 1970s; in the 1990s, CCD photometry
provided the first accurate measurements in Coma, as described by
Bernstein et al. (1995).

Two kind of problems affect these experiments: (a) the typical
surface brightness of intracluster light (ICL) is less than 1\% of the
typical sky brightness, and (b) it is difficult to disentangle between
diffuse light associated with the halo of the cD galaxy at the cluster
center and the diffuse light component.

Since 1995, wide-field cameras equipped with a CCD mosaic have allowed
accurate measurements of diffuse light in the Abell clusters.  This
stellar component is traced by tails, arcs and/or plumes with typical
$\mu_B = 27.8$ mag arcsec$^{-2}$, very narrow ($\sim 2$kpc) and
extended ($\sim 100$ kpc) in Coma and Centaurus (Gregg \& West 1998,
Threntam \& Mobasher 1998, Calc\'{a}neo-Rold\'{a}n et al. 2000).  In
addition, different groups have measured the radial surface brightness
profile of the extended cD halos out to very large cluster
radii. These measurements were carried out for the Abell cluster 1651
(Gonzalez et al. 2000), Abell 1413 \& MKW7 (Feldmeier et al. 2002) and
for the compact group HGC90 (White et al. 2003). Quoting from Uson et
al. (1991), {\it `` ...whether this diffuse light is called the cD
envelope or diffuse intergalactic light is a matter of semantics: it
is a diffuse component which is distributed with elliptical symmetry
about the center of the cluster potential''}. All these independent
measurements place the lower limit to the fraction of diffuse light in
clusters with respect to the amount of light in individual galaxies to
20\%.

\subsection{Direct detection of IC stars} 
An alternative method for probing ICL is through the direct detection
and measurements of the stars themselves.

The detection of intergalactic supernovae was first reported by Smith
(1981): a SN1a was observed in the Virgo cluster, in the region
between M86 and M84. In 2003, Gam-Yam et al. (2003) observed two SN1a
in Abell 403 ($z=0.10$) and in Abell 2122/4 ($z=0.066$) . Both events
appear projected on the halos of the central cDs, no other obvious
hosts are present, but these stars have a substantial velocity offset
(750 - 2000 km s$^{-1}$) from the cD systemic velocity suggesting that
they are not bound to it, but are free-flying in the cluster
potential. Gam-Yam and collaborators estimate that 20\% of the SN1a
parent stellar population in clusters is intergalactic.

In 1995, West and collaborators argued that a population of
intergalactic globular clusters (IGCs) exists in all clusters and are
concentrated towards the center (West et al. 1995). High values of the
GC frequency ($S_N$) in cDs is then the results of the accretion of a
number of IGCs. Independent evidence was acquired by Cot\'{e} et
al. (2001) around M87, where they concluded that the metal poor GCs
were not formed in situ, but stripped from the Virgo cluster
dwarfs. In the cluster Abell 1185, Jord\`{a}n et al. (2003) searched
for a population of IGCs, in a field centered on the peak of the
cluster's X-ray emission, which contains no bright galaxies. An excess
of point like sources is found with respect to HDF North, which
Jord\`{a}n and collaborators associate with a population of
IGCs. Bassino et al. (2003) also reported on the discovery of IGC
candidates in the Fornax cluster.
 
In the IC regions, an additional kind of stellar cluster is found.
Drinkwater et al. (2003) discovered Ultra-compact dwarfs (UDCs) in the
Fornax cluster; they are nucleated dwarf galaxies whose outer
envelopes were stripped by interactions with the cD, at the cluster center.

Direct observations of stars in Virgo IC fields were carried out by 
Ferguson, Tanvir \& von Hippel in 1998 with HST. The presence of
intracluster red giant stars (IRGBs) was inferred from the excess of
red number counts in a Virgo IC field with respect to the HDF North.
By comparing the I-band star counts into two Virgo cluster fields with
similar observations for a metal-poor nucleated dwarf elliptical,
Durrell et al. (2002) found an offset between the RGB tip of the IRGBs
and that in the dwarf. Their interpretation is that the bulk of the
IRGBs are moderately metal rich  {$(-0.8 < [Fe/H]< -0.2)$}. The
surface brightness associated with the IRGB counts is $\mu_L = 27.9$
mag arcsec$^{-2}$, and it amounts to 15\% of the Virgo cluster
galaxy I band luminosity.

Are these stars tidally stripped from galaxies during the early phases
of cluster collapse, or are they removed gradually over time via
``galaxy harassment''? Do all of these stars have parent galaxies or
do they form {\it in situ}? The recent discovery of an isolated
compact HII region in the Virgo cluster (Gerhard et al. 2002) has
shown that some star-formation activity can indeed take place in the
outskirts of galaxy halos if not already in Virgo IC space. 
 
\section{Intracluster Planetary Nebulae as tracers of cluster evolution}
Intracluster planetary nebulae (ICPNe) have several unique features
that make them ideal for probing ICL. The diffuse
envelope of a PN re-emits 15\% of the UV light
of the central star in one bright optical emission line, the green
[OIII]$\lambda 5007$ \AA\ line. PNe can therefore readily be detected in
external galaxies out to distances of 25 Mpc and their velocities can
be determined from moderate resolution $(\lambda /\Delta \lambda \sim 5000)$ 
spectra: this enables kinematical studies of the IC stellar population.

PNe trace stellar luminosity and therefore provide an estimate of
the total IC light. Also, through the [OIII] $\lambda 5007$ \AA\
planetary nebulae luminosity function (PNLF), PNe are good distance
indicators, and the observed shape of the PNLF provides information on
the line of sight distribution of the IC starlight.

ICPNe are useful tracers to study the spatial distribution,
kinematics, and metallicity of the diffuse stellar population in
nearby clusters.  Different cluster formation mechanisms predict
different spatial distributions and velocity distributions for the IC
stars.  If most of the IC light originates in the initial cluster
collapse (Merritt 1984), its distribution and kinematics should follow
closely that of galaxies in the cluster. On the other hand, if the IC
light builds slowly with time because of ``galaxy harassment'' (Moore
et al. 1996) and ``tidal stirring'' (Mayer et al. 2001) , then a
fraction of IC light may still be located in long streams along the
orbits of the parent galaxies, and
dynamically unmixed structures should be easily visible in phase
space, see the analog in the Milky way (Helmi 2001).

\begin{figure}[h]
\plotone{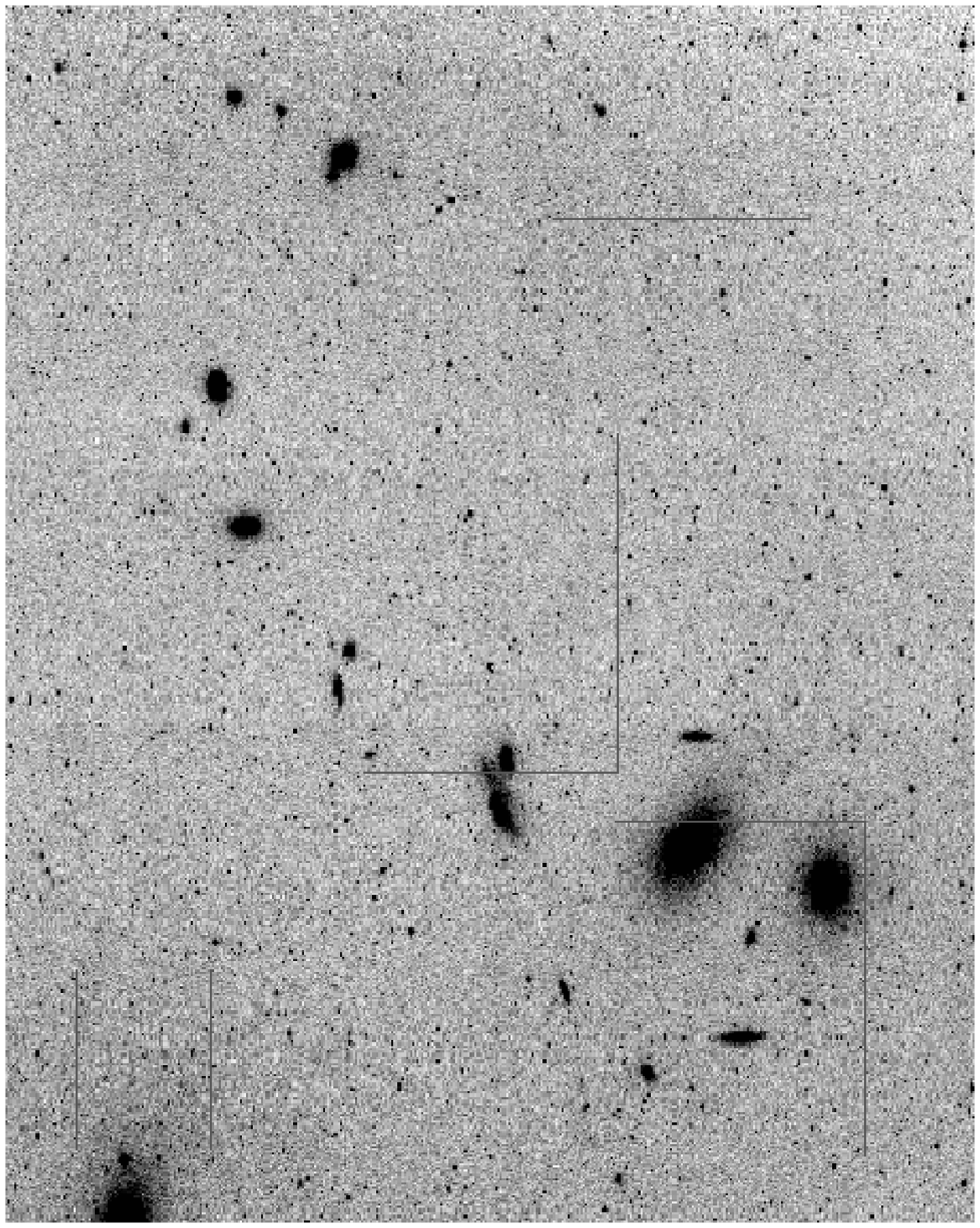}
\caption{\label{fig:VF} { \small 
Surveyed fields in the Virgo cluster: the two upper field were
obtained at the ESO MPI 2.2m telescope, and the lower-right field with the
Suprime Cam at the 8.2m Subaru telescope. The lower-left field is from
Feldmeier et al. (1998) and was used to test the selection criteria on
the spectroscopically confirmed ICPNe in Arnaboldi et
al. (2002). Several more fields need to be surveyed to determine the
large scale surface density distribution of the ICL in the Virgo
cluster.} } %
\end{figure}

\subsection{Narrow-band wide-field surveys} 
Several groups (Arnaboldi et al. 2002, 2003; Feldmeier et
al. 1998, 2003; Okamura et al. 2002) have embarked on a narrow-band
[OIII] imaging survey in the Virgo cluster, with the aim of
determining the radial density profile of the diffuse light, and
gaining information on the velocity distribution via subsequent
spectroscopic observations of the obtained samples. Given the use of
the PNLF as distance indicators, one also acquires valuable information
on the 3D shape of the Virgo cluster from these ICPN samples, see also
Feldmeier et al. (1998).

Wide-field mosaic cameras, such as the WFI on the ESO MPI 2.2m
telescope and the Suprime Cam on the Subaru 8.2m, allow us to identify
the ICPNe associated with the extended ICL (Arnaboldi et al. 2002,
2003; Okamura et al. 2002): a layout of the fields' position on the
DSS image of the Virgo core region are shown on Figure~1. These
surveys require the use of data reduction techniques suited for mosaic
images, and also the development and refining of selection criteria
based on color-magnitude diagrams (CMD) produced with SExtractor.

In Arnaboldi et al. (2002), the on-band/off-band [OIII] imaging technique which
has been used for PNe identification in Virgo and Fornax ellipticals
has been translated into the following selection criteria for the most reliable
detection of ICPN candidates:
\begin{enumerate}
\item the source should be unresolved;
\item the source should have an emission line $EW > 100$ \AA. This is
evaluated by measuring the ([OIII] - V) color between a detected object
in the on-band [OIII] image and the signal in the corresponding position
in the off-band V image. The $EW$ criterion corresponds to a 
filter-dependent color excess relative to field stars;
\item there should be no source detected in the V-band image at 
the position of the detected [OIII] source.
\end{enumerate}
The requirement on EW greatly reduces the contamination from [OII]
starburst emitters at $z\sim 0.35$. The color selection must take into
account the photometric errors in the final on-image, via 
simulation of unresolved sources.

\begin{figure}[h]
\plottwo{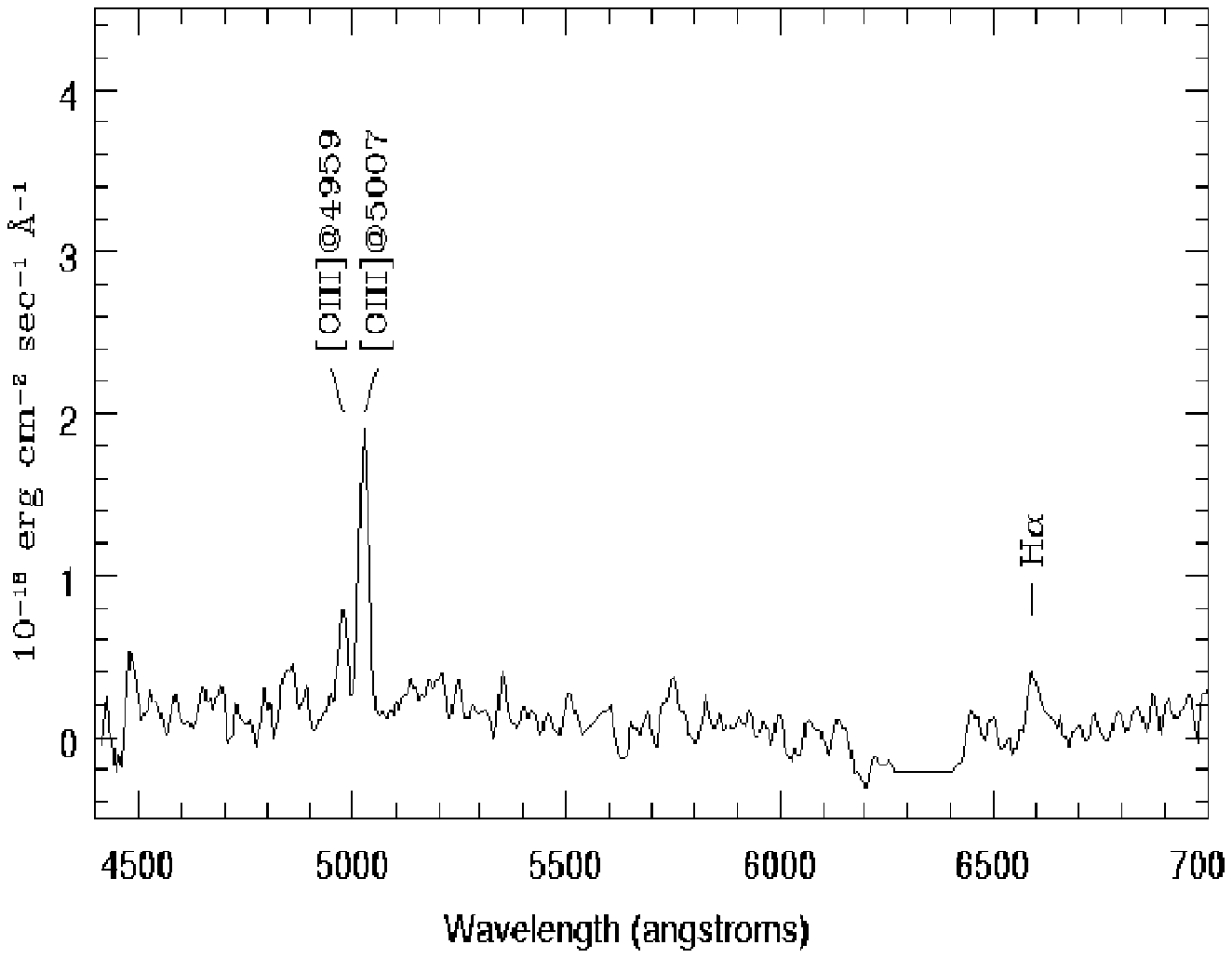}{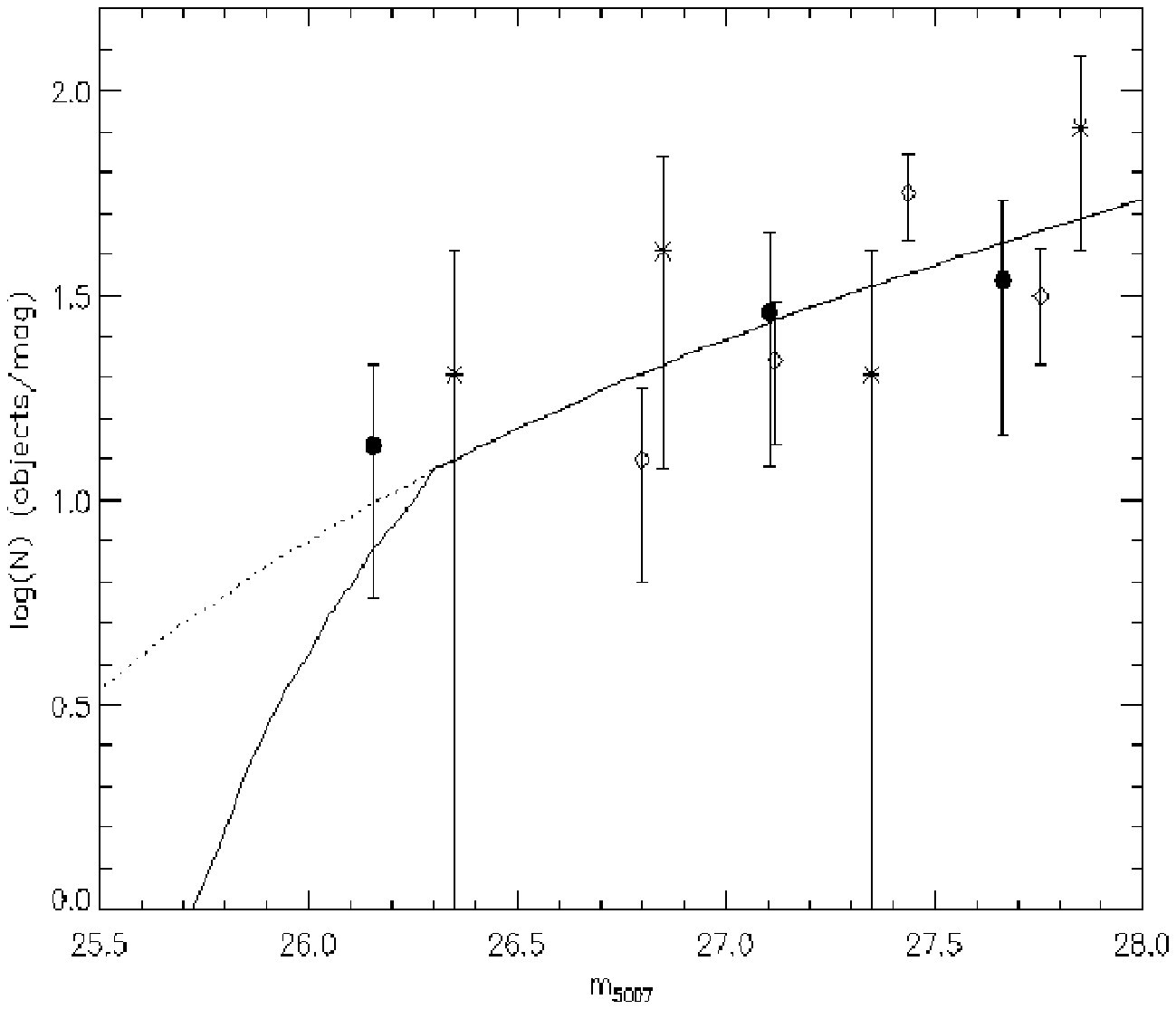}
\caption{\small Left panel $-$ Spectrum of the confirmed intracluster
PN in the Virgo cluster. The [OIII] doublet and the H$\alpha$ emission
are visible in this high S/N spectrum. Right panel $-$ The solid line
shows the expected luminosity function (LF) of the field Ly$\alpha$
population at redshift $z = 3.1$ for objects with $V < 24.73$. The
faint dotted line shows the expected Ly$\alpha$ LF without any
magnitude constraints in the V band.  Asterisks indicate the LF of
spectroscopically confirmed Ly$\alpha$ emitters from Kudritzki et al.\
(2000). Filled dots and diamonds show the LF of Ly$\alpha$ emitters in
two other blank-field surveys. These are all consistent; from
Castro-Rodriguez et al.\ (2003).}
\end{figure} 

\section{Spectroscopic confirmation and first results} 
Since the early spectroscopic detections of ICPNe by Arnaboldi et
al. (1996), the spectroscopic follow-up of the Feldmeier et al. (1998)
Virgo ICPN sample, carried out by Freeman et al. (2000) using 2dF at
the Anglo-Australian Telescope, showed that most of the emission line
sources in this sample are indeed ICPNe, because the combined spectrum
of all the ``sharp line'' emitters clearly showed the [OIII] 4959/5007
\AA\ doublet.  In 2002, a high S/N spectrum for a single ICPN in the
Virgo cluster, shown in Figure~2, was obtained for the first time at
the VLT-UT4 with FORS2 by Arnaboldi et al. (2003). We conclude that
the existence of ICPNe in the Virgo cluster is now beyond doubt.

Why then did the spectroscopic study by Kudritzki et al. (2000) find only
background galaxies?  The answer lies in examination of the luminosity
function (LF) of their objects. The LF of the candidates
studied by Kudritzki et al. (2000) follows closely the LF of field
Ly$\alpha$ emitters at $z=3.1$; see Figure~2.
 
One can compare the LF for the Ly$\alpha$ emitters with the LF for the
spectroscopically confirmed ICPNe. These confirmed ICPNe are mostly
brighter than the brightest of the Ly$\alpha$ emitters shown in
Figure~2.  The brightest of the emission line candidates
studied by Kudritzki et al. (2000) is 0.5 mag fainter than the bright
cutoff in the PNLF for M87, and 0.8 mag fainter than the bright cutoff
for the spectroscopically confirmed ICPNe in the Virgo cluster.  Most
of the current ICPN candidates in Virgo are within 1 mag of the bright
cutoff in the PNLF.  This is the reason why Kudritzki et al. did not
find ICPNe.  Their sample was dominated by the Ly$\alpha$ emitters
which are more abundant at fainter magnitudes. (See also Arnaboldi et
al. 2002).

The bright cut-off of the LF for the Virgo ICPNe is about 0.3 mag
brighter than for the PNe in individual Virgo galaxies. This is
believed to be due to the elongated structure of the Virgo cluster, as
previously found for the distribution of Virgo spiral galaxies using
the Tully-Fisher relation.

What is the fraction of Ly$\alpha$ emitters in the first magnitude of
the LF for the Virgo ICPN samples?  When Arnaboldi et al. (2002)
computed the fraction of Ly$\alpha$ emitters which can contaminate the
ICPN candidate sample selected as outlined in Section 3.1, it amounts
to about 15\% of the observed sample. This estimate is supported by
the empty field survey of Castro-Rodriguez et al. (2003).

\section{Properties of the diffuse light in Virgo cluster} 
A primary goal is to estimate the fraction of light from intracluster stars
in the surveyed region of the Virgo cluster. In our $0.25$ deg$^2$
field at a distance of $1^\circ$ from the cluster center, the ICPN
sample indicates a total associated luminosity of $ 5.8 -
7.5 \times 10^9$ L$_{B,\odot}$, which corresponds to a surface
luminosity of $ 0.33-0.57$ L$_{B,\odot}$ pc$^{-2}$ or a surface
brightness of $\mu_{B,*} = 28 - 27.7$ mag arcsec$^{-2}$. As discussed
by Arnaboldi et al. (2002), over the range of radii probed by the
survey fields, the luminosity surface density of galaxies in Virgo
decreases by a factor of $\sim 3$, while that for the ICPNe is nearly
constant. Therefore, from the data available so far, the ICPNe in Virgo
are not centrally concentrated; however we need to investigate fields at
larger radii to constrain the total amount of IC light.

One needs to compare the luminosity derived for the diffuse population
with the luminous contribution from Virgo galaxies. If ICPNe are
produced by phenomena acting locally, as the structure in the ICPN
distribution shown in Okamura et al. (2002) seems to support, then the
fraction of diffuse light with respect to the computed light in
galaxies in the field is about 10\%. On the other hand, comparing the
IC surface brightness with the smoothed out surface
brightness of galaxies from Bingelli et al. (1987) gives an upper limit 
of about 40\%.

\begin{figure}[h]
\plotone{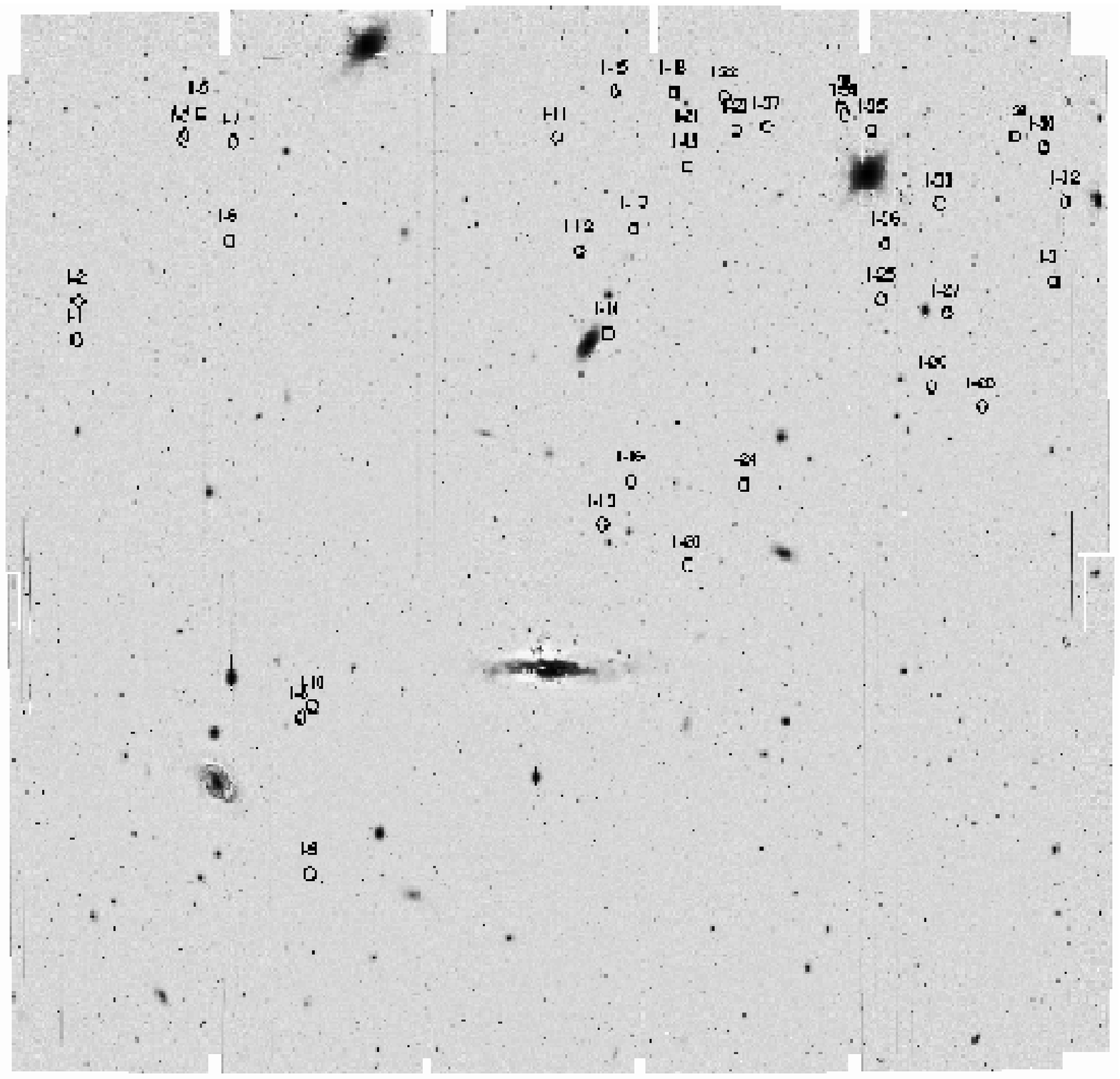}
\caption{\label{fig:ICPNdis} { \small
Deep [OIII] image of the Virgo central core region. The ICPN
candidates are marked by circles. Envelopes of bright galaxies have
been subtracted. The over-density in the upper right quadrant of this
field is highly significant. The majority of candidates seem to be
related to the M86-M84 region of the Virgo cluster, 
supporting a local origin for the ICPNe.}} 
\end{figure} 

Is the diffuse light in the Virgo cluster distributed uniformly?
Recent discoveries of low surface brightness arcs in other nearby
clusters, as discussed in Section~2, significant field-to-field
variations in the number density of Virgo ICPNe, and the remarkably
inhomogeneous distribution of ICPNe in the field surveyed by Okamura
et al. (2002), see Figure~3, have demonstrated that intracluster stars
are not distributed uniformly.

An emission line survey carried out on an empty field in the Leo group, 
using the same selection criteria as adopted for the Virgo cluster survey,
gives an upper limit on the diffuse surface luminosity of $4.4 \times 10^{-3}$ 
L$_{B,\odot}$ pc$^{-2}$, corresponding to a surface brightness limit
$\mu_{B,*} > 32.8$ mag arcsec$^{-2}$ (Castro-Rodriguez et al. 2003).  
This empty field survey, observed at the peak of the HI distribution in 
the Leo intra-group cloud, gives an upper limit on the fraction of diffuse
light in this intra group field of $ < 1.6\%$. The evidence coming
from the Leo group is very interesting because it shows that the
fraction of diffuse light vs. light in individual galaxies that we
find in Virgo is related to the Virgo cluster and its evolution. It does 
not appear to be a general physical property of the local universe.
All independent measurements carried out in the Virgo cluster IC
fields place the lower limit to the fraction of diffuse light to 10\%
of the light in individual galaxies.

\section{Intracluster stellar population properties from N-body
  cosmological simulations} 

Napolitano et al. (2003) used a high resolution simulation of a
Virgo-like cluster in a $\Lambda CDM$ cosmology to predict
the velocity and the clustering properties of the diffuse stellar
component in the intracluster region at the present epoch. The
simulated cluster builds up hierarchically and tidal interactions
between member galaxies and the cluster potential produce a diffuse
stellar component free-flying in the intracluster medium. At the end
of the simulation, the total cluster mass is $\sim 3 \times 10^{14}
M_\odot$. 0.5 million particles are within the cluster virial radius;
each particle has a mass of $0.5 \times 10^{9} M_\odot$ and the
spatial resolution is 2.5 kpc. 

Napolitano and collaborators adopt an empirical scheme to identify
tracers of the stellar component in the simulation and hence study its
properties. Stellar tracers are those particles in the simulation
within local over-densities higher than $\sim 10^4 \times \rho_{crit}$
at any given time before $z>0.25$, which is considered the epoch when
star formation stopped in the cluster. The size of an over-density
region with $12000 \times \rho_{crit}$ has a linear dimension of 15
kpc at $z=0$, and 12 kpc at $z=3$: this simple criterion selects
particles in the central regions of dark halos whose size is
comparable to galaxy luminous parts. The conversion from
stellar-mass-particle to ICPNe is based on the assumption that at
$z=0$ the luminous $M/L$ ratio of the harassed stellar matter is that of
an evolved stellar population, like that of M31, and the number of
ICPNe for a given stellar-mass are derived from an $M/L \simeq 6$ and
the luminosity-specific PN density $\alpha_{1,B} = 9.4 \times 10^{-9}
\mbox{PN\,L}_{\odot,B}^{-1}$ (Ciardullo et al. 1989).

Napolitano et al. (2003) find that at $z=0$ the ICL is mostly
dynamically unmixed and clustered in structures on scales of about 50
kpc at a radius of $400-500$ kpc from the cluster center: the
two-dimensional phase space diagrams, see Figure~4, show filaments,
cluster of particles and large empty regions, while dark matter
particles are not clustered in the same fields. The simulations
predict the radial velocity distribution expected in spectroscopic
follow-up surveys.  When they compare the spatial clustering in the
simulation with the properties of the Virgo IC stellar population, a
substantial agreement is found.

\section{Conclusions} 
Surface brightness photometry and direct detection of individual stars
give an estimate for the fraction of diffuse light in rich clusters:
it amounts to $\sim 20\%$ of the light in individual cluster galaxies.
In the nearby universe, the results obtained so far from ICPNe samples
in the Virgo cluster have shown that the fraction of the diffuse light
in the cluster amounts to 10\%-40\%, the intracluster stars are not
centrally condensed and not uniformly distributed and the front edge
of the Virgo cluster is about 20\% closer to us than M87.

A high-resolution collisionless N-body simulation of a Virgo-like
cluster at $z=0$ (Napolitano et al. 2003) predicts strong substructure
in phase-space, so the next goal will be to look for substructure in
the radial velocity distribution of ICPN candidates in Virgo.  The VLT
instruments, FLAMES and VIMOS, will be most important in giving us the
radial velocity distribution of the stars in the diffuse component,
identifying individual streams, and providing us with samples of the
phase space for the diffuse component at different cluster
radii. These observational results will be compared with N-body high
resolution cosmological simulations and in this way we should be able
to determine how old dynamically the diffuse light is.

\begin{figure}
\plotone{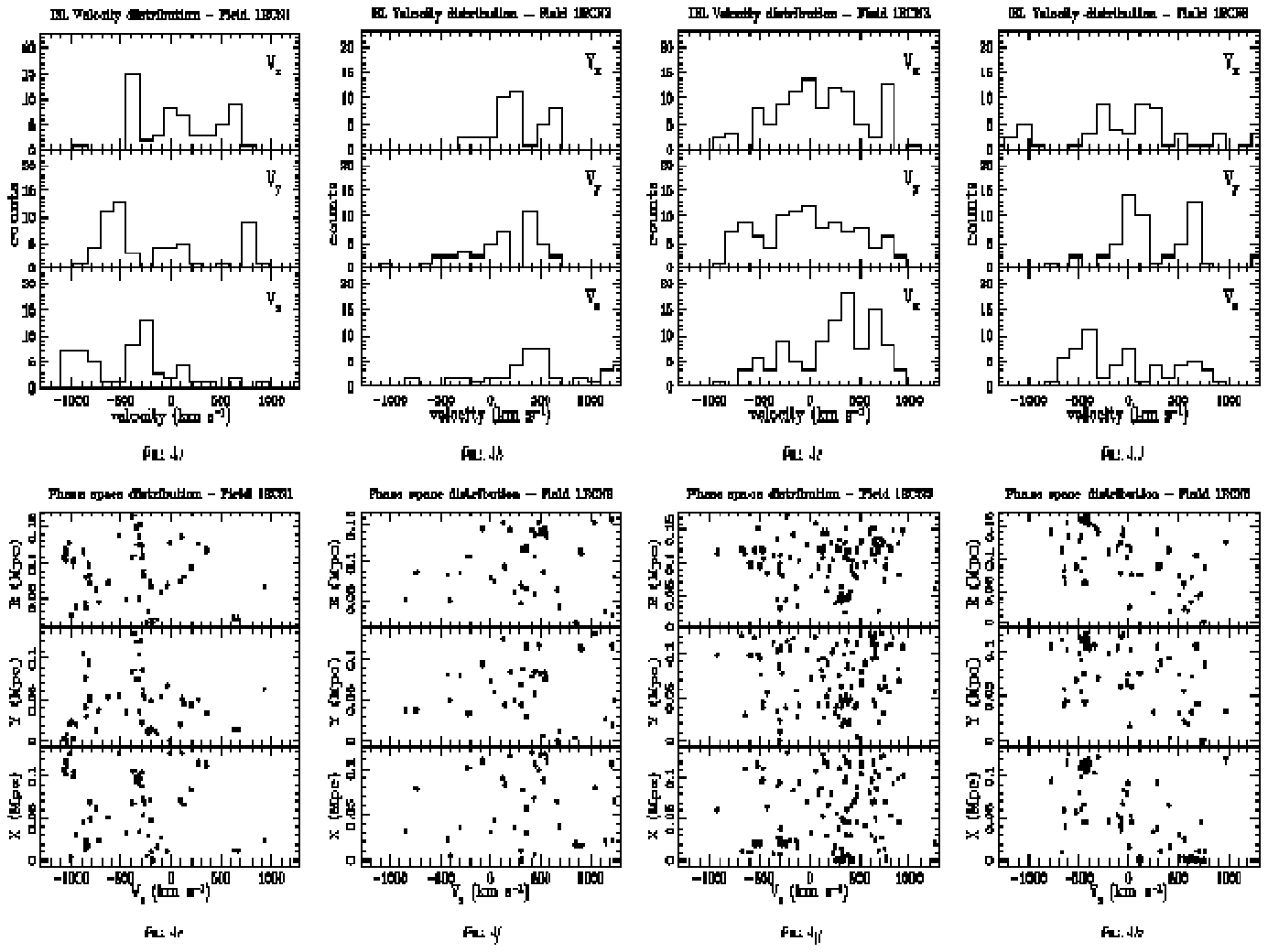}
\caption{\label{fig:phasespace} { \small 
Velocity distribution and projected phase-space diagram for the
intracluster stellar population, from Napolitano et al. (2003).}} %
\end{figure} 

\acknowledgments{ MA wishes to thank all her collaborators, without
whom these results could not have been achieved.  MA thanks the
Scientific Organizing Committee for the Invitation to give this review
at the IAU Symposium n. 217, and gratefully acknowledge the IAU for
financial support, and the INAF grant for National Projects.  }

\end{document}